\documentclass[aps,prl,twocolumn,showpacs,amsmath,amssymb]{revtex4} 
\usepackage{graphicx} 
\usepackage{dcolumn} 
\usepackage{bm} 

\begin{document}

\title{Viscous to Inertial Crossover in Liquid Drop Coalescence}

\author{Joseph D. Paulsen}
\email{paulsenj@uchicago.edu}
\author{Justin C. Burton}
\author{Sidney R. Nagel}

\affiliation{The James Franck Institute and Department of Physics, The University of Chicago, Chicago, IL 60637}

\date{\today}

\begin{abstract}

Using an electrical method and high-speed imaging we probe drop coalescence down to 10 ns after the drops touch. By varying the liquid viscosity over two decades, we conclude that at sufficiently low approach velocity where deformation is not present, the drops coalesce with an unexpectedly late crossover time between a regime dominated by viscous and one dominated by inertial effects. We argue that the late crossover, not accounted for in the theory, can be explained by an appropriate choice of length-scales present in the flow geometry. 

\end{abstract}


\maketitle

Typically, it is a simple exercise to estimate when a fluid flow will switch between a viscous and an inertially dominated regime. The dimensionless Reynolds number serves this purpose: by identifying the characteristic velocity ($U$) and length ($L$) one expects crossover behavior when $Re=\rho UL/\mu\approx1$ where $\rho$ and $\mu$ are the liquid density and viscosity respectively.  In this way, Eggers \textit{et al.} \cite{Eggers1999,Eggers2003} compute a Reynolds number for liquid drop coalescence, which predicts that for the case of salt water, viscous forces give way to inertial ones just $0.7$ ns after the drops touch. 

High-speed imaging experiments \cite{Wu2004, Thoroddsen2005, Bonn2005, Fezzaa2008} have observed the coalescence of water drops at speeds up to $10^6$ frames per second, but limited spatial and temporal resolution prevented these studies from observing the initial regime dominated by viscous effects. In this paper, we use an electrical method \cite{Case2008,Case2009} to observe salt water coalescence down to $10$ ns after the drops touch. Our measurements are conducted at slow enough approach velocities where no drop deformation occurs. Following the widely accepted Reynolds number for coalescence \cite{Eggers1999, Eggers2003, Wu2004, Yao2005, Thoroddsen2005, Bonn2005}, we would expect to see only the inertial regime. However, we observe viscous effects until $2$ $\mu$s after contact which is more than $3$ decades longer than predicted. 

We argue that the source of this discrepancy is that the previously used Reynolds number is based on an incorrect length-scale. Our detailed measurements covering two decades in liquid viscosity suggest a new picture of the crossover; the correct Reynolds number for drop coalescence is based on a length-scale, not fully appreciated previously, that reflects the dominant flows.

\begin{figure}[bt]
\centering 
\begin{center} 
\includegraphics[width=3.1in]{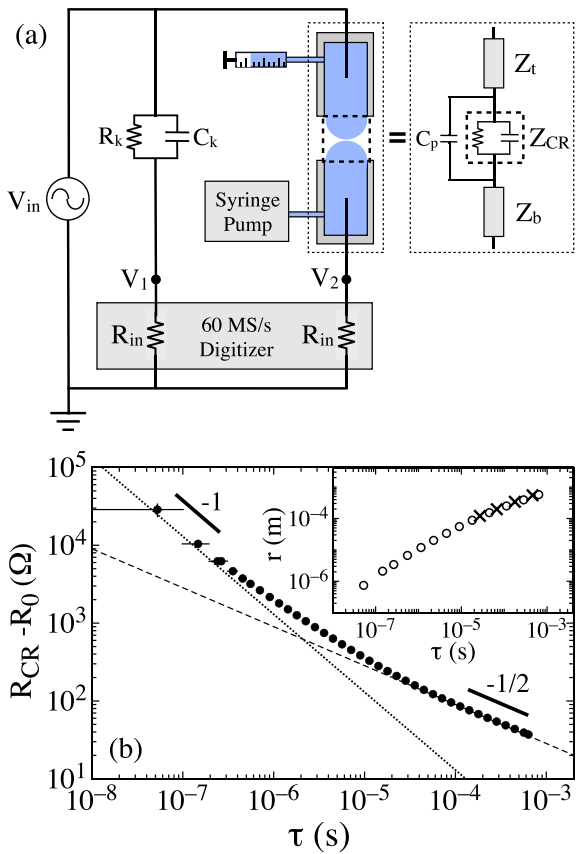} 
\end{center}
\caption{(a) Coalescence cell and measurement circuit. We apply an AC voltage, $V_{in}$ (Hewlett-Packard, HP3325A), across known circuit elements ($R_k$ and $C_k$) and coalescing liquid drops. We read voltages $V_1$ and $V_2$ into Labview (National Instruments) to calculate the cell impedance (dotted box). $Z_{CR}$ is from the coalescence region (dashed box); $Z_t$ and $Z_b$ are due to the top and bottom nozzles; $C_p$ is a small cell capacitance in parallel with $Z_{CR}$; and $R_{in}=50 \Omega$ is the input impedance of a high-speed digitizer (National Instruments, NI PCI-5105). The nozzles are brought together to measure $Z_t+Z_b$. (b) Mean value of $R_{CR}-R_0$ for six coalescences of aqueous NaCl ($\gamma=88$ mN/m, $\rho=1180$ kg/m$^3$, $\mu=1.9$ mPa s). Vertical error bars are the standard deviation of the points in each logarithmically spaced bin. Horizontal error bars are $\pm 1/(2f)$. Asymptotic behavior is consistent with $1.3\times 10^{-3} \tau^{-1}$ at early times (dotted line), and $0.90 \tau^{-1/2}$ at late times (dashed line). \textit{Inset}: Measurements of the bridge radius from the electric method ($\circ$) probe much earlier than high-speed imaging ($\times$), but extend just as far to long times.
}
\label{Schematic}
\end{figure}

\textit{Experiment}---As illustrated in Fig.\ \ref{Schematic}(a), liquid drops are formed on two vertically aligned teflon nozzles of radius $A = 2$ mm. One drop is fixed while the other drop is slowly grown with a syringe pump until they coalesce. Our experiments were performed at ambient air pressure. The intervening gas layer between two colliding drops can distort the drops and delay their coalescence \cite{Neitzel2002}. Previous experiments \cite{Case2008,Case2009} suggest that such distortion may be present for approach velocities, $U_{app}$, as low as $10^{-4}$ m/s. To this end, we have measured the effect of the ambient gas. The dynamics reported in this paper are in a low $U_{app}$ regime ($U_{app}\leq 8.0\times 10^{-5}$ m/s) where the gas does not disturb the initiation of coalescence.   

We follow the electrical method developed by Case \textit{et al.} \cite{Case2008,Case2009} to isolate the time-varying complex impedance, $Z_{CR}$, of two liquid hemispheres as they coalesce.  A high-frequency ($0.6 \leq f \leq 10$ MHz) AC voltage, $V_{in}$, is applied across the drops at low amplitude ($V_{in} \leq 1$ V). By measuring two voltages, $V_1$ and $V_2$ shown in Fig.\ \ref{Schematic}(a), we extract $Z_{CR}$, which we model as a time-varying resistor ($R_{CR}$) and capacitor ($C_{CR}$) in parallel.  By applying an additional DC offset voltage, we can determine that the electric fields do not affect the measurement of the coalescence. A sharp feature in the phase of $V_2$ at the instant the drops touch allows us to determine the moment of contact, $t_0$, to within 1/$f$. 

We calculate the conversion between $R_{CR}$ and the neck radius $r$ using the electrostatics calculation package EStat (FieldCo) \cite{Case2008, Case2009}. We compared the calculation of two possible bridge geometries while fixing the electrical potential at the nozzles. The results agree with each other.  This implies that the minimum bridge radius, $r$, is the single geometrical feature controlling the resistance. We find an excellent fit to:  
\begin{equation}
R_{CR} = 2/\xi \sigma r + R_0
\label{rtoR}
\end{equation} 
from $r=0$ out to $r=A/3$, where $\sigma$ is the electrical conductivity of the fluid, $\xi=3.62 \pm 0.05$ is a fitting parameter obtained from the simulations, and $R_0=1/\sigma \pi A$. 

According to this conversion, the measured quantity of interest is $R_{CR}-R_0$, which we show for aqueous NaCl coalescence in Fig.\ \ref{Schematic}(b) as a function of $\tau\equiv t-t_0$. In the inset, we convert this measurement to a bridge radius, $r$, and show that it agrees with high-speed imaging results. This comparison demonstrates not only the quantitative accuracy of the electric method but also its superior dynamic range as compared to the optical techniques.

\textit{Comparison to theory}---Coalescence begins in a viscous regime, where surface tension, $\gamma$, is balanced by viscous forces. Based on the analytic solution in two dimensions \cite{Hopper1990}, it is predicted that for three-dimensional drops \cite{Eggers1999}:
\begin{equation}
r_{viscous} = C_0 \frac{\gamma\tau}{\mu},
\label{viscScaling}
\end{equation}
 
\noindent where $C_0$ is calculated to be:
\begin{equation}
C_0= -\frac{1}{\pi}\ln\left({\frac{\gamma \tau}{A\mu}}\right).
\label{logcorrect}
\end{equation}

\noindent We expect $C_0$ to be nearly unity over our measurement range. High-speed imaging experiments \cite{Yao2005, Thoroddsen2005, Bonn2005} that corroborate eqn.\ \ref{viscScaling} measure prefactors of order unity, but are not sensitive to the logarithmic corrections. 

In the other limit, where inertial forces balance surface tension, a scaling argument \cite{Eggers1999} produces:
\begin{equation}
r_{inviscid} = D_0 \left( \frac{\gamma A}{\rho} \right)^{1/4} \tau^{1/2},
\label{invScaling}
\end{equation}

\noindent Numerical simulations reproduce this scaling \cite{MenchacaRocha2001, Eggers2003, Lee2006} and give $D_0\approx 1.62$ \cite{Eggers2003}. High-speed imaging experiments \cite{MenchacaRocha2001, Wu2004, Thoroddsen2005, Bonn2005, Fezzaa2008} also observe this scaling regime. 

For $2$ mm drops of aqueous NaCl solution in air, we therefore expect $R_{CR}-R_0=4.5 \times 10^{-4} \tau^{-1}$ at early times, and $R_{CR}-R_0=1.1 \tau^{-1/2}$ at late times (using eqns. \ref{rtoR}, \ref{viscScaling}, and \ref{invScaling}).  As shown in Fig.\ \ref{Schematic}(b), our measured $R_{CR}-R_0$ is in agreement with both of these predicted asymptotic scalings. 

However there is a glaring discrepancy. Theory predicts that the crossover time between these regimes, $\tau_c$, is roughly $0.7$ ns, whereas we see $\tau_c \approx 2$ $\mu$s, which is more than $3$ decades later than expected. We investigate this discrepancy by varying the liquid viscosity.

\textit{Varying the liquid viscosity}---Our liquids, mixtures of glycerol and water, were saturated with NaCl to make them electrically conductive. We measured the viscosity, surface tension, density, and electrical conductivity of each fluid. By varying the glycerol content, the viscosity could be varied over two decades (from $1.9$  mPa s to $230$  mPa s) while the surface tension and density remained nearly constant, changing by only a factor of $1.6$ and $1.04$, respectively. 

\begin{figure}[bt]
\centering 
\begin{center} 
\includegraphics[width=3.0in]{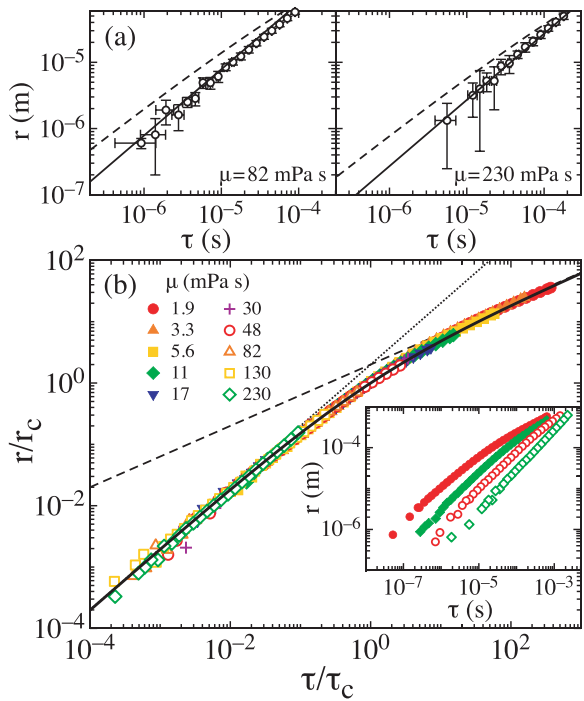} 
\end{center}
\caption{(color online) Bridge radius, $r$, versus $\tau$ for glycerol-water-NaCl mixtures of various viscosities. At each viscosity, measurements from 5 or more runs are logarithmically binned and averaged. (a) At two viscosities, data for $r(\tau)$ are compared with eqn.\ \ref{viscScaling}, with logarithmic corrections given by eqn.\ \ref{logcorrect} (dashed lines) and with $C_0=1$ (solid lines). The data fit better to $C_0=1$. (b) \textit{Inset}: $r(\tau)$ for viscosities ranging from $1.9$ to $230$ mPa s. \textit{Main}: Data are collapsed by rescaling the horizontal and vertical axes by $\tau_c$ and $r_c$ for each viscosity. Limiting behavior is proportional to $\tau/\tau_c$ (dotted line) at early times, and $\sqrt{\tau/\tau_c}$ (dashed line) at late times. The collapsed data are well described by eqn.\ \ref{interpolate} (solid line).} 
\label{rElecCollapse}
\end{figure}

The predicted viscous scaling (eqn.\ \ref{viscScaling}) with the logarithmic correction (eqn.\ \ref{logcorrect}) is an asymptotic result that has no free parameters and is predicted to apply for $r/A \lesssim 0.03$ \cite{Eggers1999}, which is difficult for optical methods to probe. Fig.\ \ref{rElecCollapse}(a) shows our data from electrical measurements that extend $2$ decades below $0.03 A$. We compare this data with $C_0$ given by eqn.\ \ref{logcorrect} (dashed lines) and with $C_0=1$, i.e., linear expansion at the capillary speed, $\gamma/\mu$ (solid lines). The data are better fit by the latter; we find no evidence for the predicted logarithmic corrections.

The inset to Fig.\ \ref{rElecCollapse}(b) shows $r(\tau)$ for ten viscosities, ranging from $1.9$  mPa s to $230$  mPa s. As shown in Fig.\ \ref{rElecCollapse}(b), the data collapses cleanly onto itself when the axes are rescaled: $\tau \rightarrow \tau/ \tau_c$ and $r \rightarrow r/r_c$ where $\tau_c$ and $r_c$ are free parameters at each viscosity to produce the best collapse. The entire data in the master curve can be well fit with the simple interpolation:
\begin{equation}
r/r_c=2 \left(\frac{1}{\tau/\tau_c} + \frac{1}{\sqrt{\tau/\tau_c}}\right)^{-1}.
\label{interpolate}
\end{equation}

\begin{figure}[bt] 
\centering 
\begin{center} 
\includegraphics[width=2.5in]{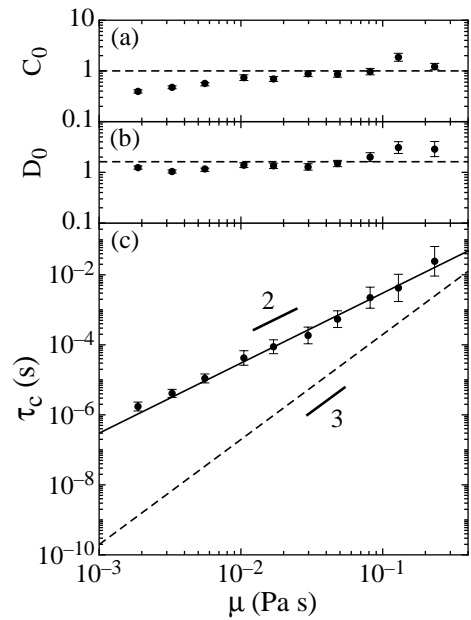} 
\end{center}
\caption{(a and b) Measured dimensionless scaling-law prefactors, $C_0$ and $D_0$, versus viscosity. In (a), the dashed line is $C_0=1$. In (b), the dashed line is $D_0=1.62$, the value obtained from simulation \cite{Eggers2003}. Error bars show the range in which $\tau_c$ and $r_c$ can be varied without affecting the quality of the collapse. (c) Viscous to inertial crossover times versus viscosity, obtained via the collapse in Fig.\ \ref{rElecCollapse}. The solid line, $\tau_c=0.3\mu^2$, is based on our proposed Reynolds number (i.e.\ eqn.\ \ref{ourtc} with $\rho=1200$ kg/m$^3$, $\gamma=65$ mN/m, $D_0=1.62$) and fits the data well. The dashed line shows $\tau_c = \mu^3/\rho \gamma^2 = 0.2 \mu^3$ obtained from the Reynolds number proposed in the literature \cite{Eggers1999, Eggers2003}. Clearly this is a poor fit to the data.} 
\label{tcvsmu}
\end{figure}

\noindent This collapse determines the coefficients for the early and late-time scaling. We compare these coefficients to the predicted values in Fig.\ \ref{tcvsmu}; our measurements of $C_0$ are of order 1 across the entire range of viscosity, and $D_0$ is in good agreement with the predicted value of $1.62$ \cite{Eggers2003}. 

At low viscosity, there is a small departure from unity in our measurement of $C_0$. A possible cause is the non-vanishing dynamic viscosity ratio between the surrounding air and our fluids. This would be larger at low liquid viscosity, consistent with the data. These effects should be negligible for $\mu \geq 82$ mPa s, the lower viscosity curve in Fig.\ \ref{rElecCollapse}(a), where the ratio is $4 \times 10^4$. We also note that if the logarithmic corrections of eqn.\ \ref{logcorrect} hold, there should be small deviations from the master curve, Fig.\ \ref{rElecCollapse}(b). These corrections are difficult to access experimentally but would be more pronounced as one goes to smaller $r$.

The crossover time, $\tau_c$, as a function of viscosity, $\mu$, is shown in Fig.\ \ref{tcvsmu}(c). The data are fit well by a quadratic dependence on $\mu$ (solid line). Clearly, the accepted formula for $\tau_c \propto \mu^3$ (dashed line) does not agree with the data. This suggests that the conventional Reynolds number for coalescence, $Re = \rho \gamma^2 \tau/\mu^3$, is wrong. This estimate is based on using $r$ and $dr/d\tau$ as the dominant length- and velocity- scales. We argue that these are not the right choices. Instead, we suggest that the dominant flows occur over a much smaller length: the neck height.

\begin{figure}[bt] 
\centering 
\begin{center} 
\includegraphics[width=2.7in]{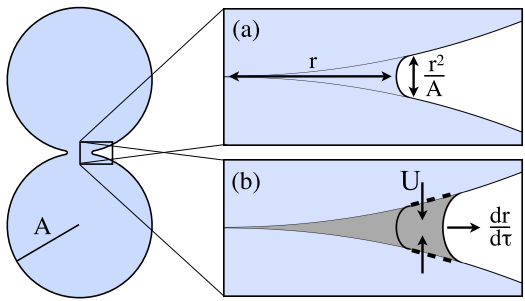} 
\end{center}
\caption{Length- and velocity- scales for coalescence. (a) In the inertial regime, a fluid bridge of radius $r$ has height $r^2/A$. (b) As the neck expands radially, fluid fills in the neighboring region with a flow, which occurs on a scale set by half the gap size, $L\approx  r^2/2A$, that is essentially axial in direction (vertical arrows). The speed of this flow, $U$, can be related to $dr/d\tau$. The volume sandwiched between the drops out to a radius $r$ (shaded region) is given by $V=\frac{\pi}{2A} r^4$. As the bridge expands, $dV/d\tau=\frac{2\pi}{A} r^3 dr/d\tau$. This fluid is supplied through two annular regions of radius $r$ and width $r^2/A$, above and below the gap (dotted lines), of total area $\frac{4\pi}{A} r^3$. Setting the volume flow rate, $\frac{4\pi}{A} r^3 U$, equal to $dV/d\tau$, we get $U\approx \frac{1}{2}dr/d\tau$.} 
\label{ReFigure}
\end{figure}

\textit{Reynolds number for coalescence}---We can equally well prescribe a Reynolds number coming from either the viscous or the inertial regime. In the inertial regime, the neck height (the spacing between the drops just outside the fluid neck) is $r^2/A$ \cite{Eggers2003}. As illustrated in Fig.\ \ref{ReFigure}, liquid from each drop moves to fill half of the gap, $L\approx r^2/2A$. We assume that the flow occurs in an annular region having a width equal to the gap size. Due to volume considerations (see Fig.\ \ref{ReFigure}), the flow speed is proportional to the interfacial velocity, $U\approx \frac{1}{2} dr/d\tau$. Using eqn.\ \ref{invScaling} for $r$, we get $L\approx \frac{1}{2} D_0^2 (\gamma/\rho A)^{1/2} \tau$ and $U\approx \frac{1}{4} D_0 (\gamma A/\rho)^{1/4} \tau^{-1/2}$.

Using these characteristic scales, the Reynolds number is: $Re=\rho U L/\mu \approx D_0^3 (\rho\gamma^3/A)^{1/4} \tau^{1/2}/8 \mu$. The crossover time, $\tau_c$, is obtained by setting $Re=1$:
\begin{equation}
\tau_c\approx \mu^2 \frac{64}{D_0^6}\sqrt{\frac{A}{\rho \gamma^3}} 
\label{ourtc}
\end{equation}
For our fluids, we find $\tau_c\approx 0.3 \mu^2$, which is in good agreement with our data (Fig.\ \ref{tcvsmu}(c)). Hence, our proposed Reynolds number gives crossover times consistent with our experiments ranging over two decades in viscosity \cite{FiveThirds}. The only place where the two predictions ($\tau_c \propto \mu^2$ versus $\tau_c \propto \mu^3$) give similar results is at high liquid viscosity, which is where two previous crossover measurements had been reported \cite{Bonn2005, Thoroddsen2005}.

This argument can be recast in terms of a crossover length. Using eqns. \ref{invScaling} and \ref{ourtc}, we find $Re \approx 1$ corresponds to when $r^2/A \approx \mu^2/\rho\gamma \equiv l_v$, where $l_v$ is the viscous length-scale. In other words, coalescence proceeds in the viscous regime until the gap between the drops becomes as large as the viscous length-scale of the fluid. Therefore, the relevant length scale to which $l_v$ should be compared is not the bridge radius, $r$ (as in \cite{Eggers1999, Eggers2003, Yao2005, Burton2007, Case2008, Case2009}), but rather the bridge height, $r^2/A$, which is much smaller. Finally, two- and three-dimensional coalescence are expected to be equivalent to leading order \cite{Eggers1999}. Thus the above argument also applies to 2D coalescence and explains an unexpectedly large crossover radius measured in the coalescence of liquid lenses \cite{Burton2007}.

\textit{Conclusion}---We have probed coalescence down to $10$ ns after the drops touch, providing a detailed study of the viscous to inertial crossover dynamics for liquid drop coalescence. The surprisingly late crossover we observe supports a new picture where a length scale drastically smaller than the bridge radius controls the flow. In the viscous regime, the data are better fit with a constant expansion velocity than the form predicted to have logarithmic corrections. One remarkable outcome is that our data, over a wide range of viscosity and time, can be rescaled onto the master curve in Fig.\ \ref{rElecCollapse}. This includes a very long crossover region between the viscous and inertial regimes, where there is no quantitative theoretical work describing the bridge dynamics.

\begin{acknowledgments} 
We are grateful to Santosh Appathurai, Osman Basaran, Sarah Case, Michelle Driscoll, Nathan Keim, Tom Witten, and Wendy Zhang for helpful discussions. This work was supported by the ICAM post-doctoral fellowship program and NSF grant DMR-0652269. Use of facilities of the University of Chicago NSF-MRSEC and the Keck Initiative for Ultrafast Imaging are gratefully acknowledged. 
\end{acknowledgments}


\begin{thebibliography}{15}
\bibitem[{Fiv()}]{FiveThirds}
\bibinfo{note}{We get the same scaling with only a slightly different prefactor
  if we use the viscous scaling (eqn.\ \ref{viscScaling}) with $C_0 \approx 1$
  (as found in our data) for $r$ and $dr/d\tau$, with the same length scale,
  $r^2/2A$. However, in that regime, the vertical scale is predicted to be
  $r^3/A^2$ \cite{Eggers1999}, which then gives $\tau_c\propto \mu^{5/3}$. This
  crossover scaling agrees equally well with our data and so we cannot rule it
  out. In this case, the data collapse in Fig.\ \ref{rElecCollapse} would only
  be approximate, and would not hold for widely different fluid parameters.}
\end{thebibliography}
\end{document}